\documentclass[10pt, a4paper, twocolumn, twoside]{article} % 10pt font size (11 and 12 also possible), A4 paper (letterpaper for US letter) and two column layout (remove for one column)

\usepackage[english]{babel} % English language hyphenation

\usepackage{microtype} % Better typography

\usepackage{amsmath,amsfonts,amsthm} % Math packages for equations

\usepackage[svgnames]{xcolor} % Enabling colors by their 'svgnames'

\usepackage[small, labelfont=bf, up]{caption} % Custom captions under/above tables and figures

\usepackage{booktabs} % Horizontal rules in tables

\usepackage{lastpage} % Used to determine the number of pages in the document (for "Page X of Total")

\usepackage{graphicx} % Required for adding images

\usepackage{enumitem} % Required for customising lists
\setlist{noitemsep} % Remove spacing between bullet/numbered list elements

\usepackage{sectsty} % Enables custom section titles
\allsectionsfont{\usefont{OT1}{phv}{b}{n}} % Change the font of all section commands (Helvetica)

\usepackage{geometry} % Required for adjusting page dimensions

\usepackage[T1]{fontenc} % Output font encoding for international characters
\usepackage[utf8]{inputenc} % Required for inputting international characters

\usepackage{XCharter} % Use the XCharter font

\usepackage{hyperref}

\usepackage{journals}

\usepackage{fancyhdr} % Needed to define custom headers/footers
\newcommand{\shorttitle}[1]{\fancyhead[CE]{\textsl{#1}}}
\newcommand{\shortauthors}[1]{\fancyhead[CO]{\textsl{#1}}}

\fancypagestyle{firstpage}{ % Page style for the first page with the title
	\fancyhf{}
        \lhead{\vspace*{0.0em} \textit{\footnotesize 22$^{\rm nd}$ European Workshop on White Dwarfs \\ S. Lauer, T. Rauch \\[-0.2em]
          August 15--19, 2022, T\"ubingen,  Germany}}
}

\date{}

\newcommand{\authorstyle}[1]{{\large\usefont{OT1}{phv}{b}{n}\color{DarkRed}#1}} % Authors style (Helvetica)

\newcommand{\institution}[1]{{\footnotesize\usefont{OT1}{phv}{m}{sl}\color{Black}#1}} % Institutions style (Helvetica)
\usepackage{titling} % Allows custom title configuration

\newcommand{\HorRule}{\color{DarkGoldenrod}\rule{\linewidth}{1pt}} % Defines the gold horizontal rule around the title

\pretitle{
	\vspace{-30pt} % Move the entire title section up
	\HorRule\vspace{10pt} % Horizontal rule before the title
	\fontsize{16}{12}\usefont{OT1}{phv}{b}{n}\selectfont % Helvetica
	\color{DarkRed} % Text colour for the title and author(s)
}

\posttitle{\par\vskip 10pt} % Whitespace under the title

\preauthor{} % Anything that will appear before \author is printed

\postauthor{ % Anything that will appear after \author is printed
	\vspace{0pt} % Space before the rule
	{\par\HorRule} % Horizontal rule after the title
	\vspace{-10pt} % Space after the title section
}

\newcommand{\newabstract}[1]{
    {\section*{Abstract}
    \bfseries #1}
  }

\usepackage[authoryear]{natbib}
\title{Newtonian pulsations of relativistic ONe-core ultra-massive DA
  white dwarfs} % The article title

\shorttitle{Pulsating relativistic ZZ Ceti stars} % The short article title for page headings

\shortauthors{C\'orsico, Althaus, Camisassa} % The short author list for page headings

\author{
  \authorstyle{A.~H.~C\'orsico,$^1$ L.~G.~Althaus,$^1$ and
    M.Camisassa,$^2$}
	\newline\newline % Space before institutions
	$^1$\institution{Facultad de Ciencias Astron\'omicas y Geof\'isicas, Universidad
          Nacional de La Plata, Paseo del Bosque s/n, (1900) La Plata, Argentina; acorsico,althaus@fcaglp.unlp.edu.ar}\\ % Institution 1
	$^2$\institution{Departament de F\'isica, Universitat Polit\'ecnica de Catalunya, c/Esteve Terrades 5, E-08860 Castelldefels, Spain}\\ % Institution 2
      }

\begin{document}

\maketitle % Print the title

\thispagestyle{firstpage} % Apply the page style for the first page

%----------------------------------------------------------------------------------------
%	ABSTRACT
%----------------------------------------------------------------------------------------

\newabstract{Ultra-massive H-rich (DA spectral type) white dwarf
  stars ($M_{\star} > 1.05 M_{\odot}$) are expected to be substantially
  crystallized by the time they reach the ZZ Ceti instability strip
  ($T_{\rm eff} \sim 12\,000$ K). Crystallization leads to a
  separation of $^{16}$O   and $^{20}$Ne (or $^{12}$C and $^{16}$O) in
  the core of ultra-massive WDs, which strongly impacts their
  pulsational properties. An additional factor to take into account
  when modeling the evolution and pulsations of WDs in this range of
  masses are the relativistic effects, which induce changes in the
  cooling times and the stellar masses derived from the effective
  temperature and surface gravity. Given the arrival of large amounts
  of photometric data from space missions such as {\it Kepler}/{\it
    K2} and {\it TESS}, it is important to assess the impact of
  General Relativity in the context of pulsations of ultra-massive ZZ
  Ceti stars. In this work, we present results of Newtonian
  gravity($g$)-mode pulsation calculations in evolutionary
  ultra-massive WD models computed in the frame of the General
  Relativity theory. }

%----------------------------------------------------------------------------------------
%	ARTICLE BODY
%----------------------------------------------------------------------------------------

\section{Introduction}

ZZ Ceti stars are pulsating DA (H-rich atmospheres) white dwarfs (WDs)
with effective temperatures in the range $10\,500 < T_{\rm eff} < 12\,600$ K.
These variable stars exhibit pulsation periods from $\sim 100$ s to
$\sim 1400$ s due to nonradial gravity($g$) modes with harmonic degree
$\ell= 1$ and $\ell= 2$ \citep{2008ARA&A..46..157W, 2008PASP..120.1043F,
2010A&ARv..18..471A, 2019A&ARv..27....7C}.
Although the vast majority of ZZ Ceti stars
are DA WDs with masses between $\sim 0.5$ and $\sim 0.8 M_{\odot}$,
$g$-mode pulsations have also been detected at least in
four ultra-massive ($M_{\star} > 1.05 M_{\odot}$)
ZZ Ceti stars so far; they are BPM~37093 \citep[$M_{\star}=
  1.1M_{\odot}$][]{1992ApJ...390L..89K},  GD~518  \citep[$M_{\star}=
  1.24M_{\odot}$][]{2013ApJ...771L...2H}, SDSS~J084021
\citep[$M_{\star}= 1.16 M_{\odot}$][]{2017MNRAS.468..239C}, and
WD~J212402 \citep[$M_{\star}= 1.16M_{\odot}$][]{2019MNRAS.486.4574R}.
It is likely that pulsating WDs even more massive ($M_{\star} > 1.30 M_{\odot}$)
will be identified in the coming years with the advent of huge volumes
of high-quality photometric data collected by space missions such as
the ongoing {\it TESS} mission \citep{2014SPIE.9143E..20R} and the
future {\it PLATO} space telescope \citep{2014ExA....38..249R}. This
big amount of data is expected to make asteroseismology a promising
tool to study the structure and chemical composition of ultra-massive
WDs \citep{2019A&ARv..27....7C}. The increasing number of
ultra-massive WDs with masses beyond $1.30 M_{\odot}$, as well as the
immediate prospect of detecting pulsating WDs with such masses, demand
new appropriate theoretical evolutionary models to analyze them.
In particular, it is necessary to calculate models that take into account
relativistic effects and to evaluate the impact of General Relativity
on the pulsational properties of ultra-massive WDs. In this exploratory
investigation, we take the first step in this direction, calculating Newtonian
pulsations on fully relativistic equilibrium models.

\begin{figure*}[t]
  \centerline{\includegraphics[width=2.00\columnwidth]{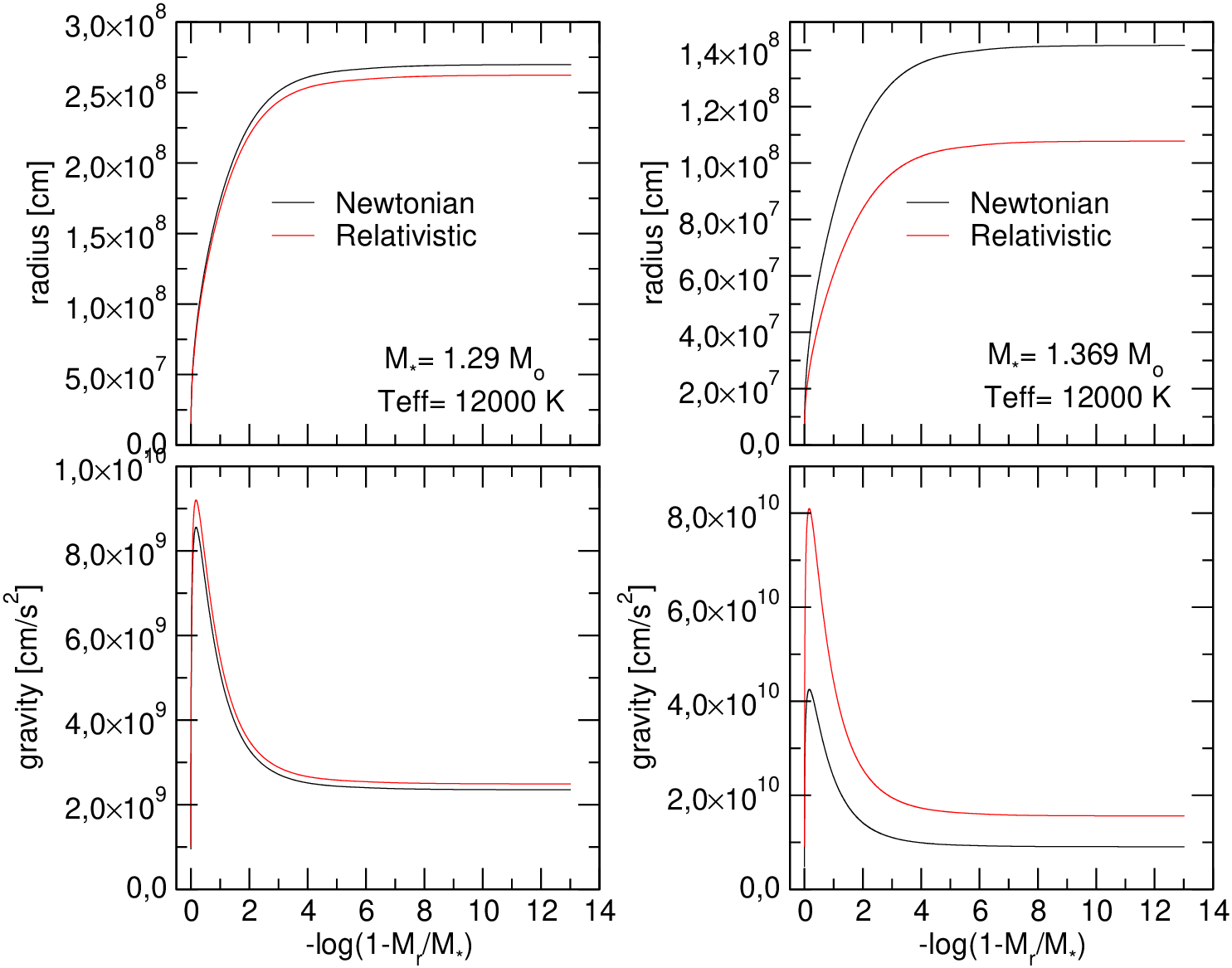}}
  \caption{The stellar radius (upper panels) and gravity (bottom panels)
    in terms of the outer mass fraction coordinate corresponding to
    ultra-massive DA WD models with $M_{\star}= 1.29 M_{\odot}$ (left) and
    $M_{\star}= 1.369 M_{\odot}$ (right), for the Newtonian case (black curves) and
    the fully relativistic case (red curves), for $T_{\rm eff} \sim  12\,000$ K.} 
  \label{fig1}
\end{figure*}

\section{Relativistic WD models}

We have generated ultra-massive WD model sequences with ONe cores
taking into account the full effects General Relativity employing the
{\tt LPCODE} stellar evolution code \citep{2022A&A...668A..58A}.
We considered realistic initial chemical
profiles as predicted by the progenitor evolutionary history
\citep{2007A&A...476..893S,2010A&A...512A..10S,2019A&A...625A..87C}, and
computed model sequences of $1.29, 1.31, 1.33, 1.35$, and $1.369
M_{\odot}$ WDs. The standard equations of stellar structure and
evolution have been generalized to include the effects of General
Relativity, following \cite{1977ApJ...212..825T}. For comparison
purposes, the same sequences have been
computed but for the Newtonian gravity case. We have included the
energy released during the crystallization process, both due to latent
heat and due to the induced chemical redistribution as in
\cite{2019A&A...625A..87C}.

We show in
Fig. \ref{fig1} the stellar radius and gravity of ultra-massive DA WD
models with $1.29 M_{\odot}$ (left
panels) and $1.369 M_{\odot}$ (right panels), for the Newtonian case
(black curves) and the fully relativistic case (red curves). Clearly,
general relativity induces smaller radii and larger gravities, and
this effect is more pronounced for larger stellar masses. 

\section{Chemical profiles and the Brunt-V\"ais\"al\"a and Lamb frequencies}

\begin{figure*}[t]
  \centerline{\includegraphics[width=2.0\columnwidth]{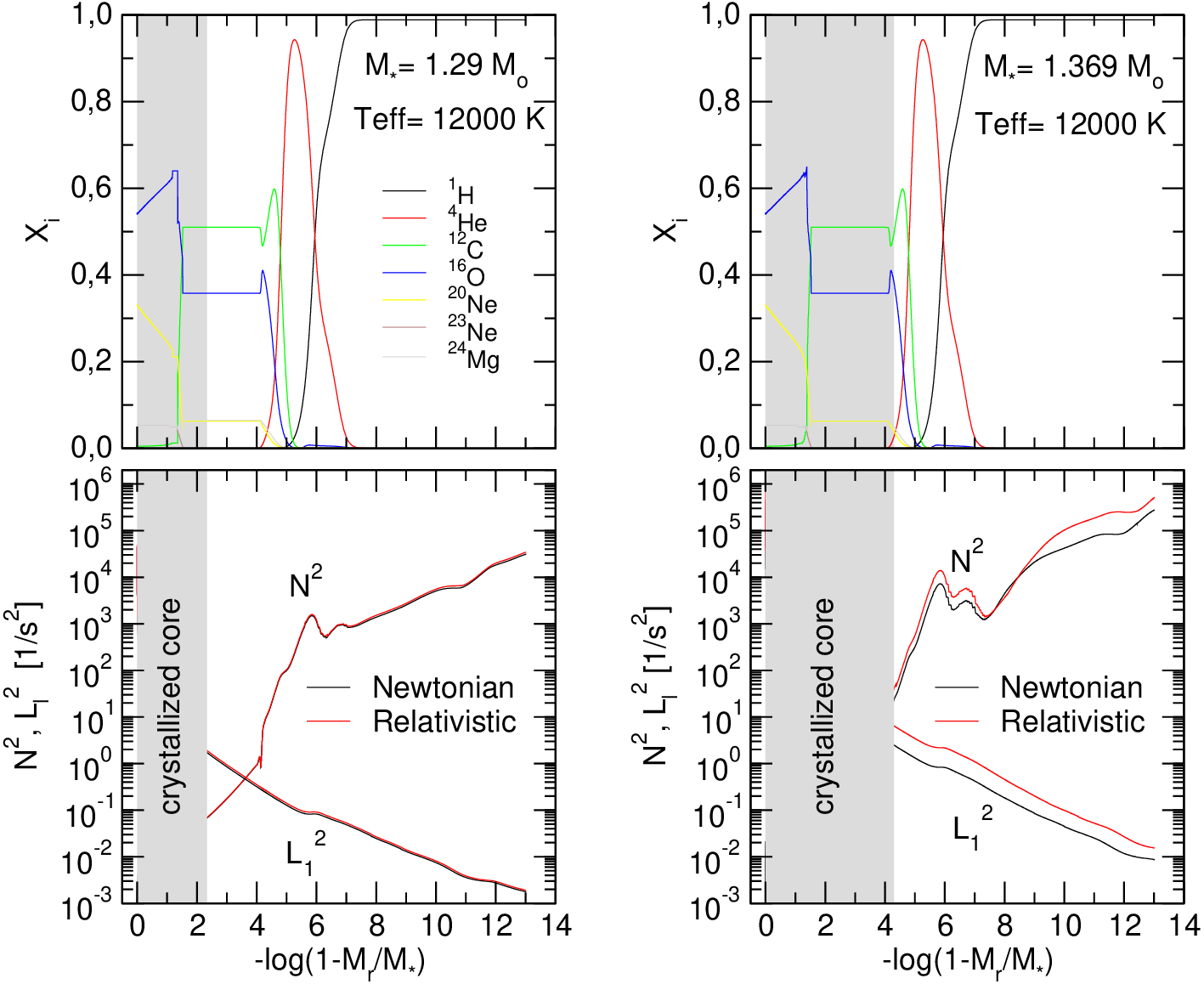}}
  \caption{Abundances by mass of the different chemical species as a
    function of the fractional mass (upper panels), and the logarithm of
    the squared Brunt-V\"ais\"al\"a and Lamb ($\ell= 1$) frequencies,
    corresponding to ONe-core DA WD models with $M_{\star}= 1.29 M_{\odot}$
    (left) and $M_{\star}= 1.369M_{\odot}$ (right), for $T_{\rm eff} \sim 12\,000$ K.
    The gray areas correspond to the crystallized regions.}
      \label{fig2}
\end{figure*}

The cores of our models are composed mostly of $^{16}$O and $^{20}$Ne and
smaller amounts of $^{12}$C, $^{23}$Na, and $^{24}$Mg. Since element diffusion
and gravitational settling operate throughout the WD evolution, our models
develop pure $^{1}$H envelopes. The $^{4}$He content of our WD sequences is
given by the evolutionary history of progenitor star, but instead, the
$^{1}$H content of our canonical envelopes [$\log(M_{\rm H}/M_{\star})= -6$] has
been set by imposing that the further evolution does not lead to $^{1}$H
thermonuclear flashes on the WD cooling track. The temporal changes of the
chemical abundances due to element diffusion are assessed by using a new
full-implicit treatment for time-dependent element diffusion
\citep{2020A&A...633A..20A}.

The chemical profiles in terms of the fractional mass for $1.29M_{\odot}$
and $1.369 M_{\odot}$  ONe-core WD models at
$T_{\rm eff} \sim 12\,000$ K and H envelope thickness
$\log(M_{\rm H}/M_{\star})= -6$ are shown in the upper panels
of Fig. \ref{fig2}.  At this effective temperature, typical of the ZZ Ceti
instability strip, the chemical rehomogeneization due to
crystallization has already finished, giving rise to a core where the
abundance of $^{16}$O increases and $^{20}$Ne decreases outward. 

In the lower panels of Fig. \ref{fig2} we show the squared
Brunt-V\"ais\"al\"a and Lamb ($\ell= 1$) frequencies corresponding to the same
models shown in the upper panels for the Newtonian case (black curves)
and the relativistic case (red red curves).  The triple chemical
transition between $^{12}$C, $^{16}$O, and $^{20}$Ne located at
$-\log(1-M_r/M_{\star}) \sim 1.5$ is within the solid part of the core,
so it has no relevance for the mode-trapping properties of the models.
This is because, according to the ``hard-sphere'' boundary conditions adopted
for the pulsations \citep{1999ApJ...526..976M}, the $g$-mode
eigenfunctions do not penetrate the crystallized region (gray areas).
In this way, the mode trapping properties
are entirely determined by the presence of the $^4$He/$^1$H transition,
which is located in more external regions. Note that the Brunt-V\"ais\"al\"a
and Lamb frequencies for the Newtonian and relativistic models are
indistinguishable for the case of $1.29 M_{\odot}$, but they notoriously differ
when $M_{\star}= 1.369 M_{\odot}$.

\section{Pulsation spectrum of $g$ modes for Newtonian and relativistic
  ultra-massive WD models}

\begin{figure}[t]
  \centerline{\includegraphics[width=1.0\columnwidth]{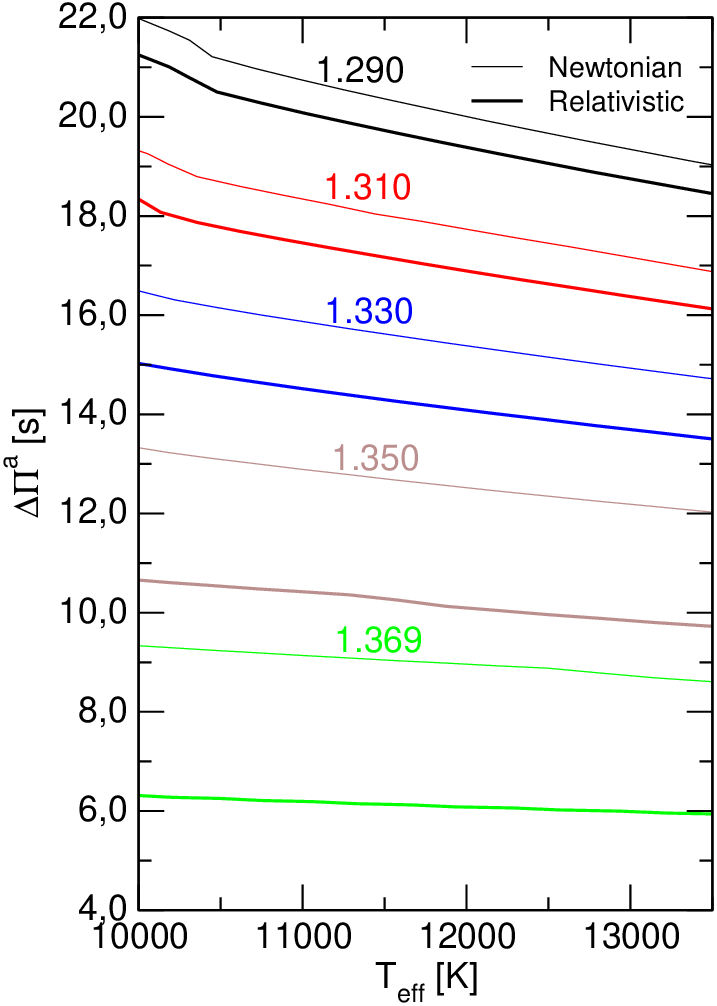}}
  \caption{The $\ell= 1$ asymptotic period spacing for WD sequences
    with different stellar masses for the Newtonian and relativistic
    cases in terms of $T_{\rm eff}$ through the whole ZZ Ceti instability
    strip.}
      \label{fig3}
\end{figure}

\begin{figure*}[t]
  \centerline{\includegraphics[width=2.0\columnwidth]{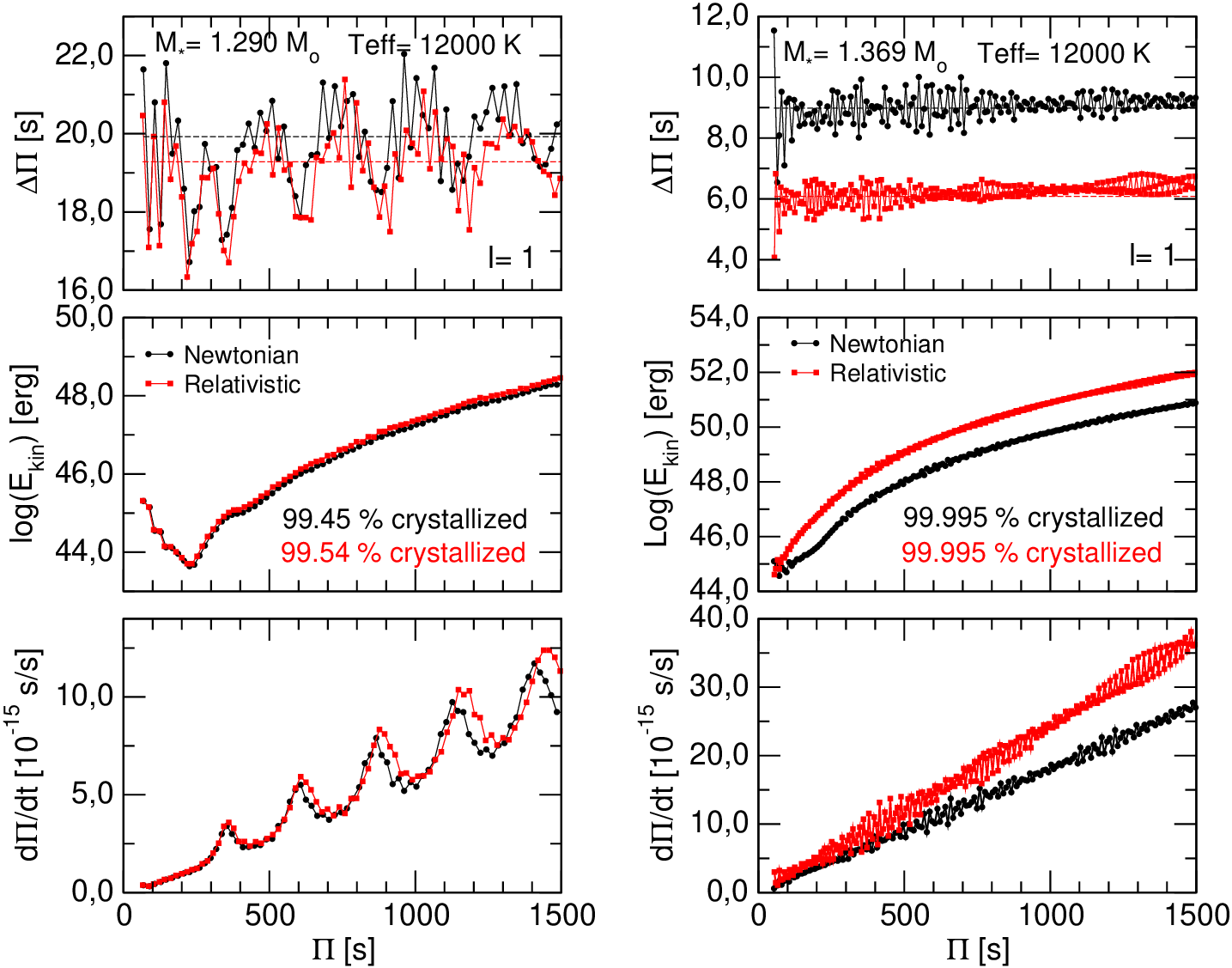}}
  \caption{The $\ell= 1$ forward period spacing ($\Delta \Pi= \Pi_{k+1}-\Pi_k$,
    upper panels),
    the kinetic energy of oscillation (middle panels), and the rate of
    period change (lower panels) for models with $1.29 M_{\odot}$ (left)
    and  $1.369 M_{\odot}$ (right) in the
    relativistic (red) and Newtonian (black) cases ($T_{\rm eff} \sim  12000$ K).}
      \label{fig4}
\end{figure*}

Adiabatic nonradial $g$-mode Newtonian pulsations have been computed
with the {\tt LP-PUL}
pulsation code \citep{2006A&A...454..863C}. This code neglects the
oscillations of $g$ modes in the crystallized region of the WD core
\citep[hard-sphere boundary condition;
  see][]{1999ApJ...526..976M,2004A&A...427..923C}.
Fig. \ref{fig3} shows
the asymptotic period spacing \citep[computed as in][]{1990ApJS...72..335T}
for the sequences
of $1.29, 1.31, 1.33, 1.35$ and $1.369 M_{\odot}$ WD models in terms of the
effective temperature all along the ZZ Ceti instability strip. We note
that the asymptotic period spacing is smaller for the case of the
relativistic WD sequences as compared with the Newtonian sequences.
This is what we expect since the asymptotic period spacing is inversely
proportional to the integral of the Brunt-V\"ais\"al\"a frequency divided
by the radius. Since the Brunt-V\"ais\"al\"a frequency is larger for the
relativistic case (see Fig. \ref{fig2}), then the integral is larger and its
inverse smaller than in the Newtonian case. The difference is larger
for larger stellar masses.

In Fig. \ref{fig4} we depict the dipole
$\ell= 1$ forward period spacing, the kinetic
oscillation energy, and the rate of period change
for the models with $1.29 M_{\odot}$  and
$1.369 M_{\odot}$, at $T_{\rm eff}= 12\,000$ K. We can see that,
in general, the period
spacing is larger in the Newtonian case, as expected (see Fig. \ref{fig3}).
On the other hand,
the oscillation kinetic energy of
the modes is higher in the relativistic case, since the WDs are more
compact and dense than in the Newtonian case. Finally, the rate of
change of periods is larger for the relativistic case, since the
cooling timescale is shorter due to relativistic effects, in
comparison with the Newtonian case  \citep{2022A&A...668A..58A}.

%------------------------------------------------

\section{Summary and conclusions}

In order to start the study of the impact of General Relativity on the
pulsations of ultra-massive WDs representative of ZZ Ceti stars, we
have calculated as an initial step the Newtonian $g$-mode pulsations
on relativistic equilibrium WD structures. We have found that the
Brunt-V\"ais\"al\"a frequency, the period
spacing, the oscillation kinetic energy, and the rates of period
change  are remarkably modified in relation to Newtonian models for
the case of very high masses, close to the Chandrasekhar mass,
although the effects are much less noticeable for lower masses. The
next step in this project is to compute the additional effects of
General Relativity on the stellar pulsations by solving the
nonradial stellar pulsation equations in the relativistic
Cowling approximation. This will allow us to assess the combined effect
of General Relativity on the equilibrium structures and on the
pulsation modes.

\section*{Acknowledgments} A.H.C. warmly thanks Klaus Werner and
the Local Organizing Committee of the  22th European White Dwarf
Workshop for support that allowed him to attend this conference.

%----------------------------------------------------------------------------------------
%	BIBLIOGRAPHY
%----------------------------------------------------------------------------------------

% There are two ways to include references. The first uses bibtex and
% is recommended. For this case uncomment the following line. 
\bibliography{papers}
% Here we have assumed that the bibliography file is named
% "papers.bib"

%----------------------------------------------------------------------------------------

\end{document}